\documentclass{article}

\usepackage{arxiv}

\usepackage[utf8]{inputenc} 
\usepackage[T1]{fontenc}    
\usepackage{hyperref}       
\usepackage{url}            
\usepackage{booktabs}       
\usepackage{amsfonts}       
\usepackage{nicefrac}       
\usepackage{microtype}      
\usepackage{lipsum}		
\usepackage{graphicx}
\usepackage{natbib}
\usepackage{doi}
\usepackage{comment} 

\usepackage{amssymb,amsmath,amsfonts,bm,dcolumn}

\renewcommand{\vec}[1]{\boldsymbol{\mathrm{#1}}}

\title{Simulation and time series analysis of responsive active Brownian particles (rABPs) with memory}

\renewcommand{\headeright}{}

\author
{
Maximilian R. Bailey,\textit{$^{a,\ast}$} Fabio Grillo, and Lucio Isa\textit{$^{a,\ast}$} 
\\
\small{$^{a}$Laboratory for Soft Materials and Interfaces, Department of Materials, ETH Z{\"u}rich, Switzerland}\\
\small{E-mail: maximilian.bailey@mat.ethz.ch}\\
\small{E-mail: lucio.isa@mat.ethz.ch}
}

\begin{document}
\maketitle

\begin{abstract}
To realise the goals of active matter at the micro- and nano-scale, the next generation of microrobots must be capable of autonomously sensing and responding to their environment to carry out pre-programmed tasks. Memory effects are proposed to have a significant effect on the dynamics of responsive robotic systems, drawing parallels to strategies used in nature across all length-scales. Inspired by the integral feedback control mechanism by which \textit{E. Coli} are proposed to sense their environment, we develop a numerical responsive active Brownian particle (rABP) model in which the rABPs continuously react to changes in the physical parameters dictated by their local environment. The resulting generated time series are then used to classify and characterise their response, leading to the identification of conditional heteroscedasticity in their physics. We then train recurrent neural networks (RNNs) capable of quantitatively describing the responsiveness of rABPs using their 2-D trajectories. We believe that our proposed strategy to determine the parameters governing the dynamics of rABPs can be applied to guide the design of microrobots with physical intelligence encoded during their fabrication.
\end{abstract}

\section{Introduction}

Inspired in part by science fiction, there is a widespread interest in miniaturising systems capable of autonomously responding to their environment to perform useful functions \cite{Bechinger2016,Ebbens2016,Tsang2020}. Despite the manifold benefits that micro- and nano-robotics could provide to applications such as biomedicine \cite{Li2017_biomed,Soto2020}, there are still significant hurdles that must first be overcome in their design and fabrication. In particular, the “intelligence" of the current state of art of synthetic microrobots - where the machine is capable of autonomously sensing the environment and then performing programmed computations to achieve human objectives - is notably lacking. This, in large part, is attributable to the difficulty in introducing sophisticated hardware at the small time-scales of interest for micro- and nanorobotics \cite{Sitti2021}. In fact, most “microrobotic" systems to date typically consist of a “microswimmers + computer" type configuration, where the tasks of sensing and adaptation are externalised - e.g. via light microscopy and particle tracking algorithms coupled to manipulation of external fields \cite{Frangipane2018,Arlt2018,Arlt2019,Koumakis2019,Khadka2018,Lavergne2019,Sprenger2020,Muinos-Landin2020,Fernandez-Rodriguez2020,Wang2023,Chen2023}

To overcome this “intelligence gap", there is a growing interest in encoding basic “physical intelligence" \cite{Sitti2021} into synthetic active materials during their fabrication. Specifically, stimuli-responsive materials such as thermoresponsive poly-NIPAM (pNIPAM) can be utilised to provide microswimmers with the ability to perceive their environment and respond in a programmed manner \cite{Alvarez2021,VanKesteren2023}. An important design parameter in such systems is the presence of “memory" \cite{Sitti2021}, which has been shown to influence the complexity of behaviours which can be achieved with responsive active systems \cite{Mijalkov2016,Leyman2018,Muinos-Landin2020,Sprenger2022}

In particular, one can draw inspiration from nature, where living systems at various length-scales, including in the micro- and nano-regime, have developed much of the computational and physical machinery desirable for autonomous applications \cite{Barkai1997,Yi2000,Vladimirov2009,Berdahl2013,Aquino2014,Lozano2019,Gosztolai2020}. Memory plays an important role in the ability of micro-organisms to perform “useful" tasks such as painting \cite{Frangipane2018,Arlt2018,Koumakis2019}, and memory has more generally been proposed to be a fundamental aspect of molecular machinery, enhancing the ability of bacteria to sense \cite{Berg1977,Aquino2014} and chemotaxis \cite{Berg1975,Yi2000,Skoge2014,Gosztolai2020}. To date, most studies with synthetic/numerical systems have relied on discrete time delays to encode memory \cite{Fernandez-Rodriguez2020,Muinos-Landin2020,Holubec2021,Wang2023}, which, while effective, relies on a feedback system controlled by an external computer. However, encoding such capacity via physical intelligence is highly challenging. Nevertheless, the manner by which micro-organisms encode memory, for example via integral control type systems with an effectively delayed response \cite{Yi2000,Aquino2014,Gosztolai2020}, could provide instructive for the design of responsive microrobots. 

In the current societal context, one cannot discuss the role of intelligence without also referencing the looming specter of artificial intelligence (AI). With respect to micro- and nanorobotic systems, AI has found application in particle tracking \cite{Midtvedt2021,Midtvedt2022a,Bailey2022,Pineda2023}, estimating parameters of their physics \cite{Bo2019,Argun2020,Argun2021,Gentili2021,Munoz-Gil2021,Ruiz-Garcia2022}, developing swimming strategies \cite{Tsang2020a,Qin2023}, and determining optimal navigation paths \cite{Muinos-Landin2020,Nasiri2022,Mo2023,Tovey2023}. In particular, “deep learning" strategies excel at extracting underlying representations of complex data - otherwise intractable using standard statistical tools - due to their intrinsic ability to model hierarchies and non-linearities present within the datasets. As dynamical systems, studying micro- and nanorobots often involves the analysis of their generated time series data to obtain a greater understanding of their physical properties. Of the various deep-learning architectures, recurrent neural networks (RNNs) are often the best suited to this task due to their ability to capture sequential dependencies and thus account for the “history" or "memory" of the data \cite{lipton2015critical}. This property has been exploited in several recent studies to determine the governing physics of a range of micro-scale systems with significant success \cite{Bo2019,Argun2020,Gentili2021,Munoz-Gil2021}.

To assist the design of micro- and nanorobotic systems which can adapt to the complex environment through which they locomote, we develop a numerical scheme of responsive active Brownian particles (rABPs) whose continuous response to their 2-D environment intrinsically imparts memory of their internal system onto their dynamics. Specifically, we build upon a previous numerical model where the rABP responds via an aphysical zero-order-hold (ZOH) delay to a checkerboard environment \cite{Fernandez-Rodriguez2020}, to motion through a double sinusoid, where the response of the rABP is governed by an integral-control type feedback inspired by the mechanisms which govern chemotaxis in bacterial systems \cite{Barkai1997,Yi2000}. Moving from a discrete feedback in a discretised environment to a continuous response in a differentiable environment not only represents a more realistic system, but also generates time series data which can be used for statistical analysis and to train RNNs which extract the physics governing the response of our rABPs to their local environment. Our numerical simulation and time series analysis therefore provides a framework by which RNNs can be trained to extract physical parameters from trajectories of real experimental systems, in turn providing insights into the responsiveness and suitability of different materials to impart physical intelligence onto active materials depending on the application desired.


\section{Numerical method}

Our numerical model builds upon the previous system developed in \cite{Fernandez-Rodriguez2020}. In brief, we simulate the dynamics of the responsive active Brownian particles (rABPs) by solving the following equations of motion:

\begin{align}
	m\ddot{x} =& \, f_x(\vec{r},\tau,\theta) - \gamma_T\dot{x} + \sqrt{2k_BT\gamma_T}\eta_x(t) \nonumber\\ 
	m\ddot{y} =& \, f_y(\vec{r},\tau,\theta) - \gamma_T\dot{y} + \sqrt{2k_BT\gamma_T}\eta_y(t) \nonumber\\ 
	I\ddot{\theta} =& \, \gamma_R(\vec{r},\tau)\dot{\theta} + \sqrt{2k_BT\gamma_R(\vec{r},\tau)}\eta_\theta(t)
	\label{eqn:langevin}
\end{align}

where $m$ and $I$ are the mass and the moment of inertia of the rABPs (density $\rho = 2500$ kg/m$^3$ and $R = 0.9e^{-6} \mu$m, for correspondence to typical experimental realisation of active colloids). $f_x(\vec{r},\tau,\theta)$ and $f_x(\vec{r},\tau,\theta)$ are the x and y components of the active force acting on the rABPs, but in contrast to \cite{Fernandez-Rodriguez2020}, also has a dependence on the response variable $\tau$ and the xy- coordinates of the particles due to the positional dependence of its velocity $V(\vec{r},\tau)$. $f_x(\vec{r},\tau,\theta)$ and $f_x(\vec{r},\tau,\theta)$ are thus given by $V(\vec{r},\tau)\text{cos($\theta$)}\gamma_T$ and $V(\vec{r},\tau)\text{sin($\theta$)}\gamma_T$ respectively, where $\gamma_T$ is the friction coefficient for translational motion. $\gamma_R(\vec{r},\tau)$ is the state-dependent friction coefficient for the rABPs' rotational dynamics, while $\eta_x(t)$, $\eta_y(t)$, and $\eta_\theta(t)$ are uncorrelated random numbers satisfying the conditions:

\begin{align}
    \langle\eta_x\rangle =& \,\langle\eta_y\rangle = \,\langle\eta_\theta\rangle = 0 
    \nonumber \\  
    \langle\eta_x^2\rangle =& \,\langle\eta_y^2\rangle = \,\langle\eta_\theta^2\rangle = 1
    \label{eqn:noise}
\end{align}

As in \cite{Fernandez-Rodriguez2020}, we solve Eq. \ref{eqn:langevin} in the underdamped limit for faster convergence,  using a Verlet-type integration scheme proposed by Gronbech-Jensen and Farago applying the It$\hat{\text{o}}$ convention \cite{Gronbech2013a}. 

\begin{figure}
    \centering
    \includegraphics[width=0.95\linewidth]{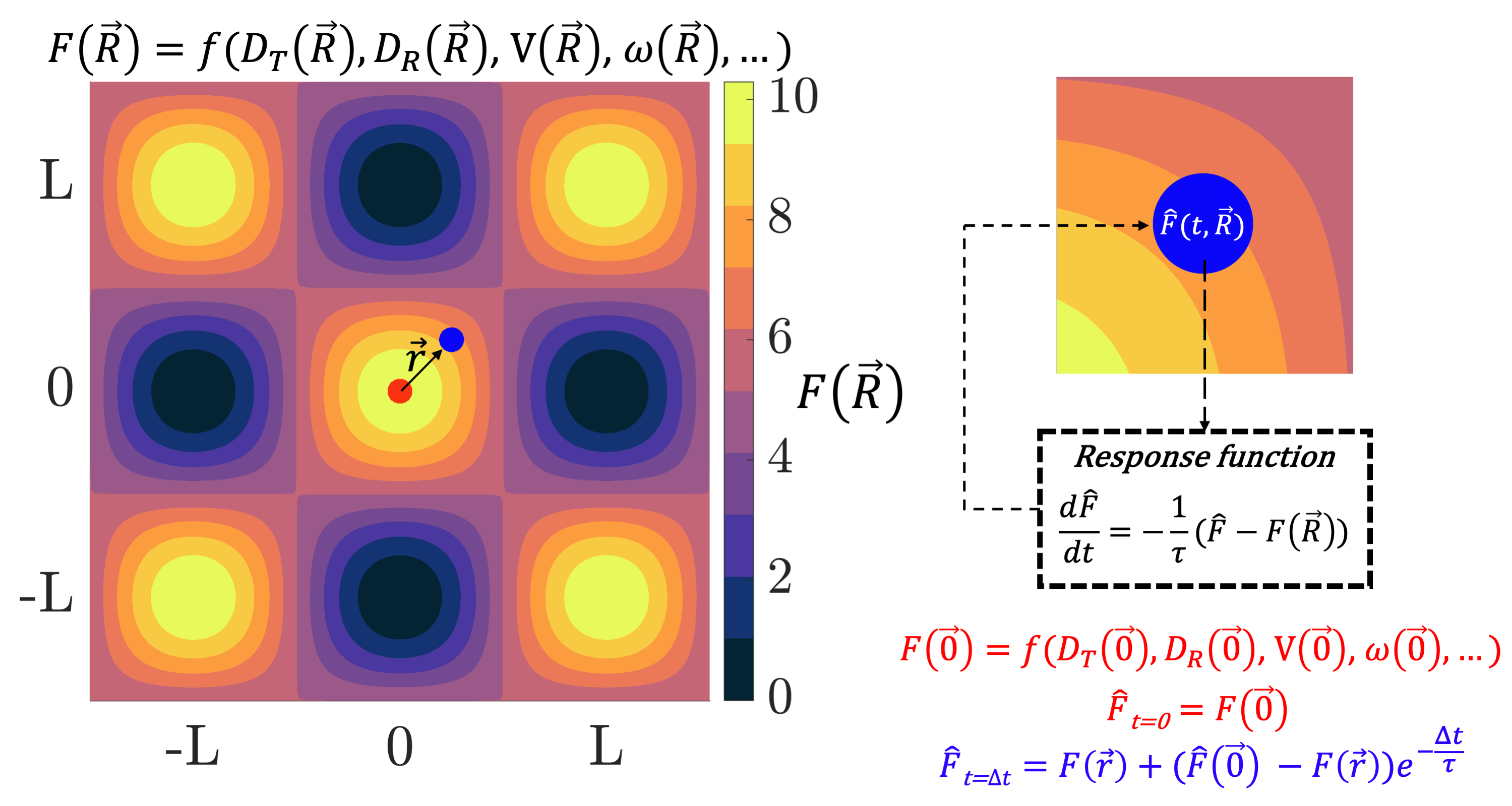}
  \caption{Schematic of the numerical simulations of a responsive active Brownian particle (rABP) moving in a spatially fluctuating environment of some variable $F(\overrightarrow{R})$. The particle is initialised at the origin at $t = 0$ (red disc), and moves during a simulation time-step $\Delta t$ to it's new position $\overrightarrow{r}$ (blue disc). At $t=0$, the particle's dynamics $\hat{F}_{t=0}$ are determined completely by its position $F(\overrightarrow{0})$. However, the update of the internal particle dynamics at $t=\Delta t$, $\hat{F}_{t=\Delta t}$, is non-Markovian and governed by the response function $d\hat{F}/dt$.}
  \label{fig:Fig_Schematic}
\end{figure}

The value of the position-dependent rotational diffusion coefficient $D_{R}$ ($D_R(\vec{r}) = k_BT/\gamma_R(\vec{r})$) and its velocity $V$ are set by the rABPs spatial position $\vec{r} = [x(\text{t}),y(\text{t})]$, via the following double sinusoidal mapping:

\begin{align}
F(\vec{r}) =& \frac{F_{min}-F_{max}}{2}\left[ 1 + \text{sin}\left(\frac{\pi x}{L}+\frac{\pi}{2}\right)\text{sin}\left(\frac{\pi y}{L}+\frac{\pi}{2}\right)\right] + F_{max}
\label{eqn:mapping}
\end{align}

where $F$ is the general form of the physical quantity determined by the spatially varying environment (e.g. $D_R(\vec{r})$, $V(\vec{r})$ - see Figure \ref{fig:Fig_Schematic}), and $F_{min}$ and $F_{max}$ determine the range of values that the mapping can take.

The value input into the rABP's governing equations of motion Eq. \ref{eqn:langevin} at time t, $\hat{F}_{t=\text{t}}$, is determined by both its position $\vec{r}$ via $F_{t=\text{t}}(\vec{r})$, and by its “memory" of its previous values $\hat{F}_{t=\text{t}-\Delta\text{t}}$. The importance of the rABPs history to its physics is thus set by $\tau$, which determines the responsiveness of the moving rABP to $F(\vec{r})$ via the response function $\frac{d\hat{F}}{dt}$ (see Figure \ref{fig:Fig_Schematic} and the main text for more details). As with the environment, the response function $\frac{d\hat{F}}{dt}$ can take any form, however we have studied the simplest form of an integral-control feedback system, given by:

\begin{align}
\frac{d\hat{F}}{dt} =& -\frac{1}{\tau}\left(\hat{F} - F(\vec{r})\right)
\label{eqn:response}
\end{align}

where $\tau$ determines the rate at which the current value of the rABP, $\hat{F}$, is able to “match" the value set by its spatial position $F(\vec{r})$. The time-scale of the simulations is set via $L/V$, the simulation time taken for a ballistically moving particle to travel the distance of one half period of the sinusoidal mapping (for example, $L/V = 10$ represents 10 simulated seconds). The simulation is incremented in 0.001 $\Delta t$ time steps, excepting for $\tau = 1e-3$ - where $\Delta t =$ 0.0001 s, while the trajectories are sub-sampled as in experiment.

\section{Results \& Discussion}
\subsection{Localisation of a responsive active Brownian particle (rABP) in a double sinusoidal mapping of $D_R$}

\begin{figure}
\centering
\includegraphics[width=0.75\linewidth]{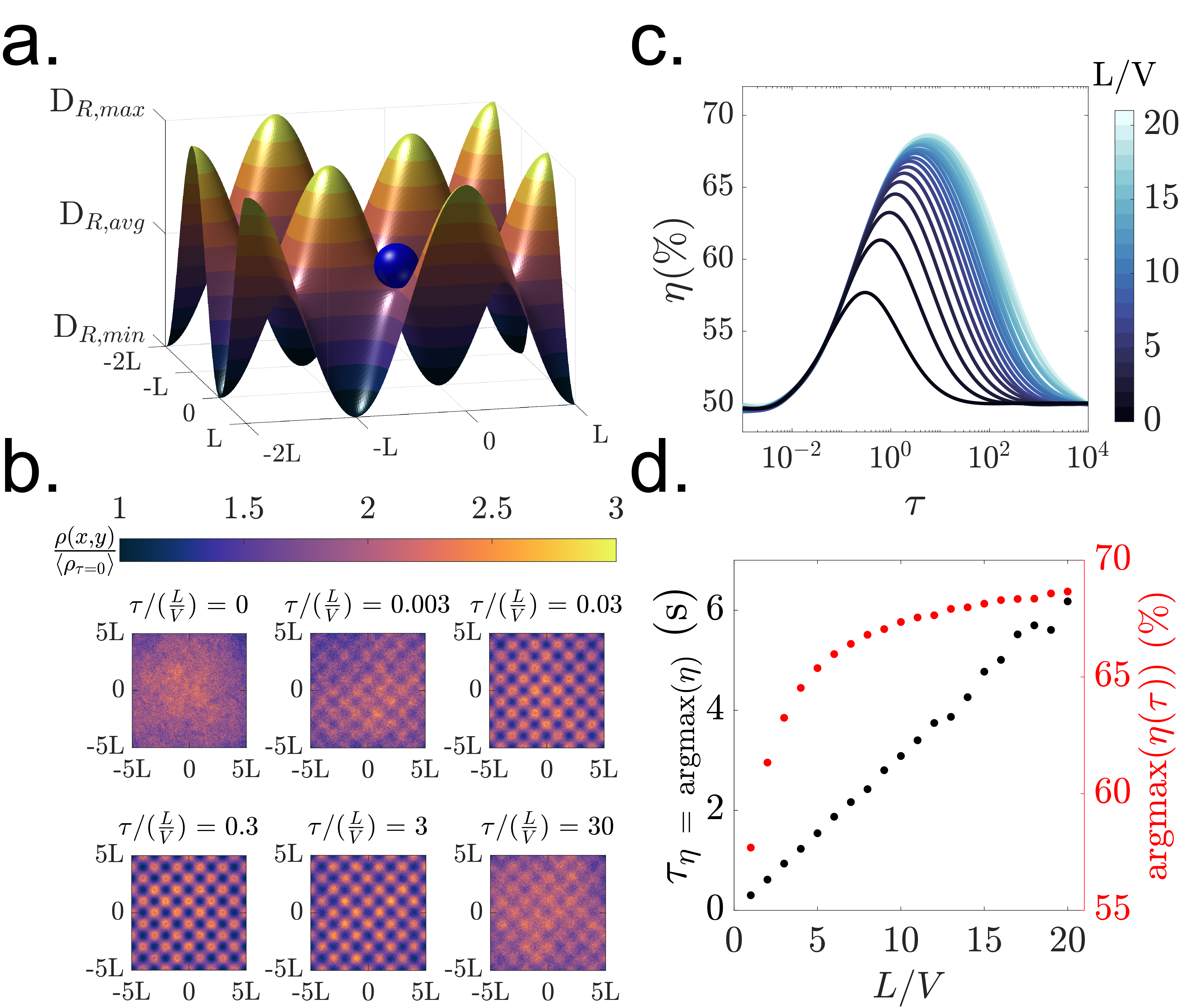}
\caption{Localisation of rABPs in a spatially varying environment where $D_R(\vec{r})$ varies between 0.01 and 10 according to a double sinusoidal mapping (in x and y). a) Schematic representation of an rABP (blue sphere) moving in the spatially varying $D_R(\vec{r})$, ranging from values of $D_{R,min}$ to $D_{R,max}$ with period $2L$. b) Varying the extent of localisation in high $D_R$ regions ($D_{R,H}$) with increasing response variable $\tau$ for a given environment time-scale of fluctuation $L/V = 10$. At intermediate response times $\tau/(L/\langle V\rangle) = 0.3$, localisation is maximised in $D_{R,H}$ regions, as illustrated by a heat map of particle positions. For small and large response times, the extent of localisation is significantly reduced, indicating the existence of an optimal $\tau$. c) Extent of particle localisation $\eta$ in $D_{R,H}$ regions, defined as $D_R(\vec{r}) \geq (D_{R,max}+D_{R,min})/2$, as a function of response time for different system time-scales $L/V$. d) The response time at which localisation is optimised, $\tau_{\eta = \text{argmax}(\eta)}$, increases linearly with the system time-scale $L/V$, while the maximum localisation $\eta_{max}$ begins to saturate around $L/V = 10$.}
  \label{fig:Fig1}
\end{figure}

We begin our study by evaluating the localisation of 1000 non-interacting rABPs moving through an environment characterised by a spatial variation in the set rotational diffusion coefficient, $D_{R}(\vec{r})$, where the mapping is given by Eq. \ref{eqn:mapping}. As in \cite{Fernandez-Rodriguez2020}, we vary the values of $D_R$ via 3 orders of magnitude from $D_{R,L} \approx 0.01$ to $D_{R,H} \approx 10$ (see Figure \ref{fig:Fig1}a). However, here we implement a continuous integral-control type feedback to govern the response of the rABP to its environment, rather than the discrete time-feedback loop previously studied. In this case, the value of $\tau$ governs the rate $d\hat{D}_R/dt$ of the response to changes in the external environment, rather than defining a discrete-time after which the value of $\hat{D}_R$ is updated. With respect to Eq. \ref{eqn:response}, the value $\hat{D}_R - D_R(\vec{r})$ can be thought as the error between the system target state $D_R(\vec{r})$ and its current state $\hat{D}_R$, similar to integral feedback control, a response observed in a range of biological systems including those governing bacterial chemotaxis \cite{Barkai1997,Yi2000}. 

We find that rABPs continuously responding to a differentiable environment will display qualitatively similar localisation behaviour as that previously reported for such particles with a discrete-time response to a checkerboard pattern \cite{Fernandez-Rodriguez2020}. Specifically, for different $L/V$ landscapes, where $V$ here is a constant at $5 \mu$ms$^{-1}$, we note the existence of an optimal response time $\tau$ for which the localisation of rABPs in $D_{R,H}$ regions, $\eta = N_H/(N_L+N_H)$ - where $N_L$ and $N_H$ are the number of rABPs in the $D_{R,L}$ and $D_{R,H}$ regions respectively - is maximised (see Figure \ref{fig:Fig1}b). Very low values of $\tau$ correspond to a near-instantaneous update of $\hat{D}_R$, and therefore no localisation is observed as previously described in \cite{Fernandez-Rodriguez2020}. Likewise, at very high values of $\tau$ the rABP is effectively “blind" to the spatial variations in $D_R(\vec{r})$, and the distribution reflects the original initialisation of the particles in their environment. 

The reduction in argmax($\eta$) compared to \cite{Fernandez-Rodriguez2020} is largely attributed to the smooth gradient in $D_R(\vec{r})$, which, coupled to the gradual response of the rABPs to their environment (rather than an instantaneous update after a discrete delay-time), provides the particles with a greater probability of escaping a high $D_R$ ($D_{R,H}$) region into a low $D_R$ one ($D_{R,L}$). More specifically, particles cannot ballistically enter $D_{R,H}$ regions in the absence of a ZOH-delay type feedback response \cite{Fernandez-Rodriguez2020}, as their $\hat{D}_R$ value will continuously rise, increasing the likelihood that they are trapped diffusively near the edges, or are able to orient away with intermediate $\hat{D}_R$ values and escape into $D_{R,L}$ regions. This is also reflected in the heat maps of the particle position (see Figure \ref{fig:Fig1}c) for low and intermediate $\tau/(L/V)$, where we observe the absence of particles in the central parts of the $D_{R,H}$ regions, indicating the inability of particles to penetrate deeply within the high $D_R$ regions.

Notably, as $\tau$ increases further beyond the optimal localisation time (e.g. $\tau/\frac{L}{V} > 0.3$), this distinction is blurred as the rABPs no longer update sufficiently quickly to sense the spatial variations in $D_R(\vec{r})$ and thus are able to travel further into the $D_{R,H}$ regions before re-orienting (see Figure \ref{fig:Fig1}c, $\tau/(L/V) \geq 3$). As a further result of the continuous update, and unlike the system studied in \cite{Fernandez-Rodriguez2020}, the maximum value of $\eta$ attained is not a linear function of $L/V$ but instead saturates around $L/V = 10$ (see Figure \ref{fig:Fig1}d). This again arises due to the continuous mapping in $D_R$, preventing the distinct delineation between “diffusive" ($D_{R,H}$) and “ballistic" ($D_{R,L}$) regions at steady state, as previously argued in \cite{Fernandez-Rodriguez2020}. Nevertheless, we note that the lag time at which $\eta$ is maximised, $\tau_{\eta=\text{argmax}(\eta)}$, is linear with $L/V$, which we attribute to the need for rABPs to be less responsive to their environment to ensure that they penetrate sufficiently far into the $D_{R,H}$ domain, and thus minimise the probability of re-orienting and escaping into a $D_{R,L}$ region.

\subsection{Statistical analysis of rABP time series with auto-regressive models}

\begin{figure}
\centering
  \includegraphics[width=0.95\linewidth]{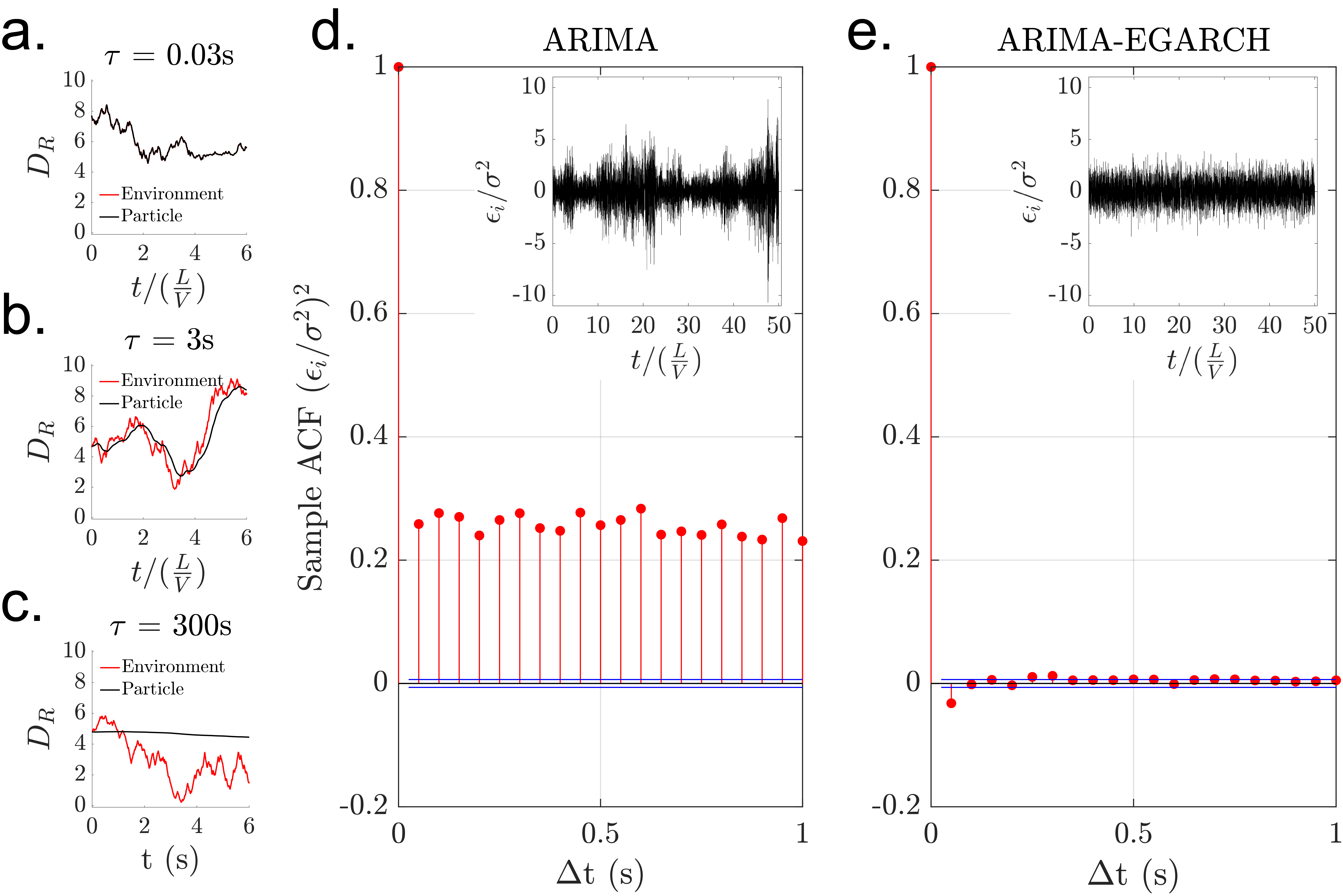}
  \caption{The continuous response of the simulated rABP - contrasting with the discrete zero-order-hold (ZOH) delay investigated in \cite{Fernandez-Rodriguez2020} - allows the characterisation of its dynamics using time series analysis techniques. a-c) As $\tau$ increases, the ability of the particle to update it's internal $\hat{D}_R$ term to follow $D_R(\vec{r})$ decreases, and there is a smoothing and delay effect reminiscent of an integral feedback controller. d) The auto-correlation function (ACF) of the normalised squared-residuals after fitting an auto-regressive ARIMA(5,2,0) model to the time series $\hat{D}_{R,\tau=3}$ indicates the presence of conditional heteroscedasticity in the time series of $\hat{D}_{R,\tau=3}$. Inset: the time series of the normalised residuals. e) Both the Akaike and Bayesian information criteria (AIC, BIC respectively) indicate that $\hat{D}_{R,\tau=3}$ is best modelled as an ARIMA(5,2,0)-EGARCH(1,1) process amongst the conditional heteroscedasticity models tested. The previously detected auto-correlation of the normalised squared-residuals when fitting only an ARIMA model also disappears. Inset: the time series of the normalised residuals. Note the absence of large fluctuations, contrasting to the case where the conditional heteroscedasticity was not accounted for. Here, $L/V = 10.$
  }
  \label{fig:Fig2}
\end{figure}

As a consequence of studying rABPs which now continuously respond to a differentiable spatial mapping of their physical parameters, we now obtain two additional time series', 1) the particle's “set" value by the environment, $F(\vec{r})$ (here $D_R(\vec{r})$), and 2) the value experienced by the particle as a function of its memory, $\hat{F}_{t=\text{t}}$ (here $\hat{D}_R$). We see clearly the effect of increasing $\tau$ on the ability of the particle to respond to its environment in Figure \ref{fig:Fig2}a, which progressively smooths and delays the response $\hat{D}_R$ to the environment value $D_R(\vec{r})$, drawing parallels to an integral-feedback controller \cite{Barkai1997,Yi2000}.  

To study the generated time series, we first turn to a frequently invoked, parsimonious family of parametric models used to describe stationary time series and make short-term forecasting predictions, namely derivatives of the auto-regressive-integrated-moving-average (ARIMA) process \cite{MAKRIDAKIS1997}. Such models approximate a time series as a combination of its $p$ previous values and/or $q$ error values ($p$ AR and $q$ MA terms respectively, see Eq. \ref{eqn:ARIMA}), while the $I$ refers to the number of differencing steps required to obtain a stationary process. Specifically, we concentrate on the time series of $D_R(\vec{r})$ experienced by the rABP for $\tau/(L/V) = 0.3$, ($\hat{D}_{R,\tau=3}$), previously identified as the value for which $\eta$ is near its maximum. 

\begin{align}
X_t =& \, \epsilon_t + \sum_{i=1}^{p}\phi_i X_{t-i} + \sum_{j=1}^{q}\theta_j \epsilon_{t-j} 
\label{eqn:ARIMA}
\end{align}

where $\epsilon_t$ represents the white-noise error (or innovation) at time $t$, $\phi_i$ denotes the $i^{th}$ auto-regressive parameter relating the value of time series at $t$, $X_t$, to its previous value $X_{t-i}$, and $\theta_j$ is the $j^{th}$ fitting parameter of the moving-average (MA) process describing the dependence of the current value $X_t$ on its error at time $t-j$, $\epsilon_{t-j}$.

Preliminary investigations showed that standard ARIMA models were not capable of adequately capturing the dynamics of $\hat{D}_{R}$, with the best fits obtained using a ARIMA($p$=5, $I$=2, $q$=0) model for $\hat{D}_{R,\tau=3}$ displaying relatively weak time dependency on its previous values. Importantly, the squared standardised residuals of the model showed significant signs of autocorrelation (see Figure \ref{fig:Fig2}d), a key marker for conditional heterocedasticity \cite{Tse2002}. Conditional heteroscedasticity refers to situations where the non-constant variance observed is itself a function of previous variance terms. Common examples are to be found in econometrics, e.g. the case of volatility clustering, with a common example being that shocks in a share price are more likely to be followed by further volatility (high variance). The most widespread approach to treat conditional heteroscedasticity in a time series is to use the autoregressive-conditional-heteroscedasticity (ARCH) family of models \cite{Engle1982}. From comparing the Akaike information criterion (AIC) of different generalised-ARCH (GARCH) models, the exponential-GARCH (EGARCH) model was found to provide the best results, and removed the presence of autocorrelation in the squared standardised residuals (see Figure \ref{fig:Fig2}e). The EGARCH model extends GARCH processes by allowing for asymmetric effects in the volatility of a time series, a feature commonly found in the stock market for which the model was proposed (shares are expected to be more volatile after bad news than good news \cite{Nelson1991}). In the context of our rABPs, the classification of $\hat{D}_R$ as an ARIMA-EGARCH process can be rationalised by considering the asymmetric variance with which a ballistic (low $D_R$) and diffusive (high $D_R$) will change their spatial co-ordinates and thus internal $\hat{D}_R$ value (large $\Delta D_R(\vec{r})$, low $\Delta D_R(\vec{r})$ respectively). Although providing some useful insight into the dynamics of an rABP's response, such models are nevertheless inadequate for a more quantitative description of its motion-determining parameters. 

\subsection{Characterising the response of rABPs with recurrent neural networks (RNNs)}
\subsubsection{Predicting $\tau$ for rABPS moving in a double sinusoidal mapping of $D_R$}

In the last few years, recurrent neural networks (RNNs) of various forms have demonstrated significant successes in the quantification of anomalously diffusing processes \cite{Bo2019,Argun2021,Gentili2021,Munoz-Gil2021}. We are therefore motivated by the question of whether such networks can also be utilised to extract meaningful information on the underlying physics governing the motion of our rABPs. Specifically, the responsiveness of the rABPs to their spatially varying environment is determined by $\tau$ (see Eq. \ref{eqn:response}), which in turn dictates the importance of their history and thus their overall dynamics (see Figure \ref{fig:Fig1}). We note that in Figures \ref{fig:Fig2}a-c, a clear increase in the smoothing-and-delaying of $\hat{D}_R$ can be observed as a function of $\tau$. We thus hypothesise that it is possible to extract information on $\tau$ by examining the differences between $F_{t=\text{t}}(\vec{r})$ and $\hat{F}_{t=\text{t}}$ (beginning with the case of $D_R$). Specifically, we investigate the ability of gated-recurrent-units (GRUs) \cite{cho2014properties,chung2014empirical} - a specific type of RNNs which performs especially well at capturing short-term dependencies in time series (as identified with the ARIMA-GARCH family of models) - to estimate $\tau$ from the available time series.

To generate sufficient training data for the recurrent neural networks, we simulate 10 rABPs for 748 equally log-spaced $\tau$ values between 0.01 and 10 (i.e. spanning 3 orders of magnitude) for $L/V = 10$, noting that the number of trajectories used is orders of magnitude less than that studied in previous works \cite{Bo2019}. We begin by sub-sampling our simulated trajectories every 50 time steps for 1200 data points, corresponding to an video acquisition rate of 20 frames-per-second (FPS) for a 60s active matter experiment. The raw input data of particle $i$ consists of the time series $D_{R,i}(\vec{r})$ and $\hat{D}_{R,i}$, while the label for regression is $\tau_i$, which governs the response of the rABP to its environment. As the values for $D_R$ and $\tau$ take positive values spanning 3 orders of magnitude, we take the common logarithm (log$_{10}$) before differencing the time series inputs to remove underlying trends, and then re-scale the values to take near integer values. We then reshape the trajectories from $[N\times 2] \rightarrow [N/2\times 4]$, i.e. treating 2 sequential time steps as a single data point, which we found to improve the model fitting \cite{Argun2020}. 
To construct and train our deep learning architecture, we make use of the \textit{keras} API \cite{chollet2015keras}, which has a range of in-built features suitable for network training. Our sequential network consists of 2 pre-processing layers of multilayer perceptrons (MLPs), which feed into 8 GRU layers, both of which have 32 hidden units, before a final regression layer with a single output to estimate $\hat{\tau}$. We use the in-built \textit{Adam} optimiser with a learning rate of 0.001, noting that we performed minimal hyper-parameter tuning (as we were not overly interested with obtaining the smallest loss possible). We trained our network with a 70:15:15 train:validation:test split, using the mean-squared error on the log-transformed labels as the performance metric. To integrate our pre-processing steps with the data input into the network, we used a Deeptrack 2.0 \cite{Midtvedt2021} generator function to continuously sample and feed the training data into our network architecture batch-wise. 

\begin{figure}
\centering
  \includegraphics[width=0.95\linewidth]{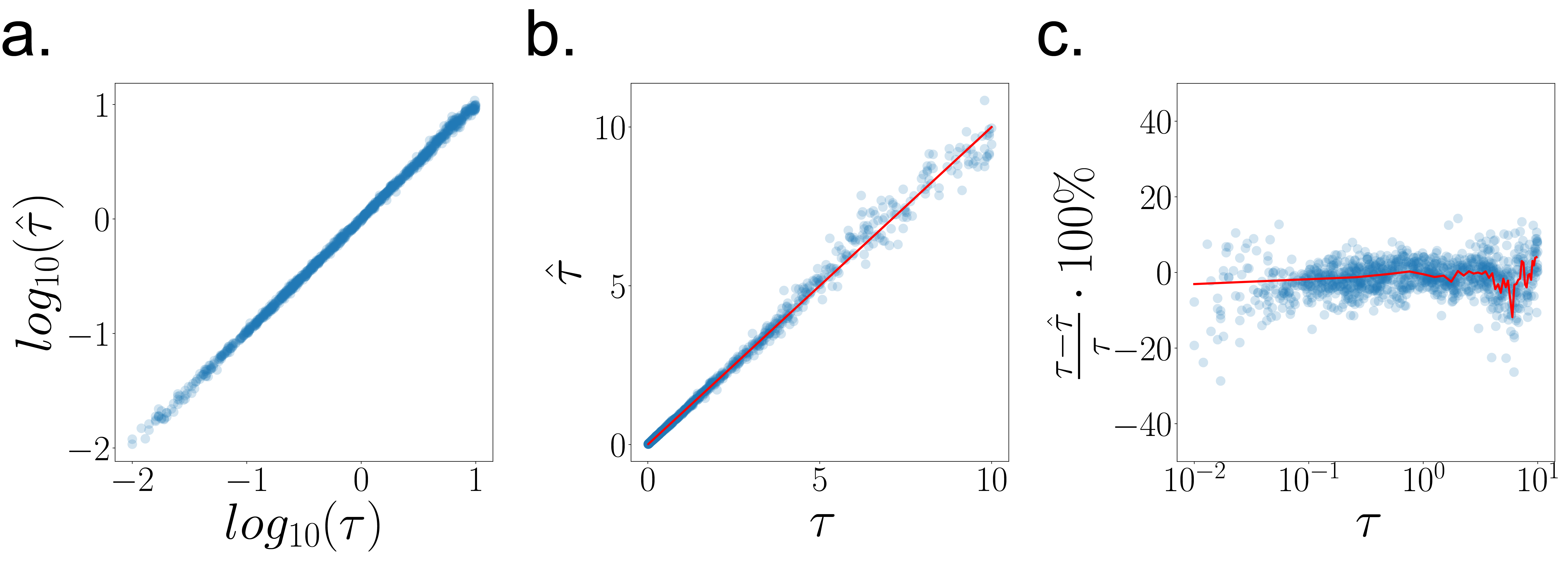}
  \caption{Fitting the response value $\tau_i$ of different rABPs from their true rotational diffusion $\hat{D}_{R,i}$ and the value set by their position $D_{R,i}(\vec{r})$ using recurrent neural networks (RNNs) where the system time-scale is given by $L/\langle V \rangle = 10$s. a) The values of $\tau$ span 3 orders of magnitude, and we therefore apply a $log_{10}$ transform to the labels before fitting. b) Reverting the fitted values $\hat{\tau}$ and $\tau$ to their original scale. c) We observe little structure in the normalised residuals of the fits, noting that at the highest and lowest $\tau$ values studied, the network performs worse. 748 different $\tau$ values with 10 different particle realisations for each $\tau$ were simulated to obtain the data for model training and evaluation.}
  \label{fig:Fig3}
\end{figure}

Given the relative simplicity of our pre-processing steps and network architecture, as well as the little hyperparameter tuning performed, the ability of our network to predict $\tau$ from the provided time series of $D_R(\vec{r})$ and $\hat{D}_R$ is impressive (see Figure \ref{fig:Fig3}). Across the range of $\tau$ values investigated, the network estimates $\hat{\tau}$ mostly within 10\% of the true value (see Figure \ref{fig:Fig3}c), with minimal structure in the residuals. We note that for the lowest and highest values of $\tau$, the network struggles more, which reflects the almost instantaneous changes in $\hat{D}_R$ for very small $\tau$ (where the two time series almost completely overlap), and the nearly imperceptible differences in $\hat{D}_R$ for large $\tau$. In particular, the performance at the lowest $\tau$ values is particularly surprising given the sampling rate which is in fact larger than the smallest values of $\tau$ studied ($\Delta t = 0.05$s, while $\tau = 0.01$s).


\subsubsection{Predicting $\tau$ for rABPS moving in a double sinusoidal mapping of $V$}

The fitting procedure outlined here relies on both $D_{R,i}(\vec{r})$ and $\hat{D}_{R,i}$ to be accessible parameters to predict $\tau$ for the $i^{\text{th}}$ rABP. $D_{R,i}(\vec{r})$ can be readily extracted from experiment if the spatial mapping of the spatially varying variable is known, e.g. due to its external imposition on the particles \cite{Fernandez-Rodriguez2020}. In contrast, $\hat{D}_{R,i}$ cannot be easily determined from particle coordinates as for the environmental set-point. This lies in the difficulty of obtaining accurate estimates of $D_R$ from ABP dynamics, which is typically extracted from their mean-squared-displacement (MSD) \cite{Bailey2022_PRE}. In particular, the problem of poor statistics and small $\Delta D_R$ is likely to be a significant challenge when attempting to determine a dynamic $\hat{D}_{R}$ from particle trajectories. A notable exception may be the case of the “raspberry" active particles studied in \cite{Niggel2023}, where the rotations can be explicitly determined, and therefore the value of $\hat{D}_R$ could be evaluated at shorter time-scales. Nevertheless, it remains questionable whether the approach outlined above using RNNs to estimate $\tau$ from $D_{R,i}(\vec{r})$ and $\hat{D}_{R,i}$ would find broader application in systems currently studied.

Nevertheless, we do expect that the approach outlined here could be used for the spatial mapping of certain particle properties. For example, the translational velocity of active particles can be readily extracted from their trajectories, without requiring the fitting of the full ABP MSD equation. Indeed, most experimental systems with an externally imposed variation in the dynamics of active materials typically focus on modulating their speed, often via light patterns \cite{Lozano2016,Lozano2019,Frangipane2018,Arlt2018,Arlt2019,Koumakis2019,Caprini2022}. Motivated by these considerations, we therefore evaluate the ability of our numerical model to capture the dynamics of rABPs locomoting through a double sinusoidal mapping of $V$, and the ability of RNNs to extract details of their underlying response. For physically relevant values, we modulate the velocity of the rABPs from $V_{min} = 1 \mu$ms$^{-1}$ to $V_{max} = 10 \mu$ms$^{-1}$, corresponding to active Péclet numbers in the range [7.6, 75.8], where $Pe = V/\sqrt{D_TD_R}$. 

\begin{figure}
\centering
  \includegraphics[width=0.95\linewidth]{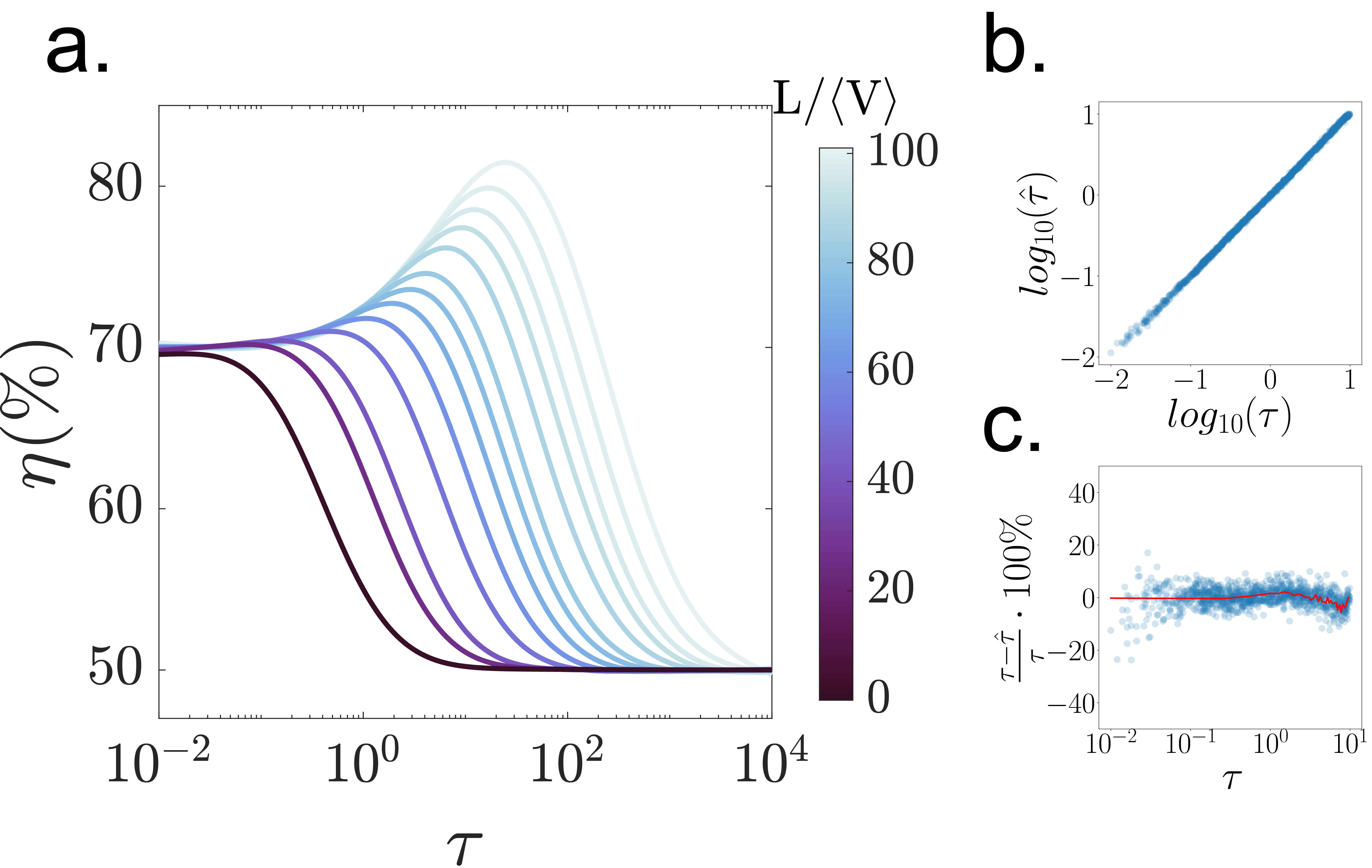}
  \caption{Implementing velocity $V$ as the response parameter of the rABPs, which varies between $V_{min} = 1 \mu ms^{-1}$ and $V_{max} = 10 \mu$ms$^{-1}$. a) Particle localisation $\eta$ in low $V(\vec{r})$ ($V_L$)  regions, defined as $V(\vec{r}) < (V_{max}+V_{min})/2$, as a function of response time for different system time-scales $L/\langle V\rangle$. b) Fitting the response time $\tau$ from their true particle velocity $\hat{V}$ and the value set by their position $V(\vec{r})$ using the gated recurrent unit (GRU) network architecture developed for $D_R$,  where the system time-scale is again given by $L/\langle V \rangle = 10$s. c) As in Figure \ref{fig:Fig3}c, we observe minimal structure in the normalised residuals of the fits, with poorer network performance for the lowest $\tau$ values. 748 different $\tau$ values with 10 different particle realisations for each $\tau$ were simulated to obtain the data for model training and evaluation.}
  \label{fig:Fig4}
\end{figure}

Unlike the systems with a spatial mapping of $D_R$, we find that even with instantaneous localisation ($\tau \rightarrow 0$) we observe some degree of localisation of rABPs in low velocity regions $V_L < \langle V\rangle$ (see Figure \ref{fig:Fig4}a). This can be rationalised by considering the relation $V_L\rho_L = V_H\rho_H = C$ at steady state, where $C$ is some constant, a property of out-of-equilibrium systems (forbidden to those at thermal equilibrium) theoretically predicted for a range of systems \cite{Schnitzer1993,Tailleur2008,Stenhammar2016,Caprini2022} and demonstrated with swimming bacteria by Arlt et al. \cite{Arlt2019}. Considering the mean velocities in the $V_L$ and $V_H$ regions (3.25 $\mu$ms$^{-1}$ and 7.75 $\mu$ms$^{-1}$ respectively), from this relation we expect a localisation $\eta \approx 70\%$ which is indeed what we observe (see Figure \ref{fig:Fig4}a). Otherwise, we again note a similar behaviour as observed for $D_R$, with a non-monotonic relationship between $\tau$ and $\eta$ as a function of $L/\langle V\rangle$. In particular, at high $\tau$ we once more recover an equal distribution in the $V_L$ and $V_H$ regions as the particles are unable to respond to changes in the environment. Interestingly, we note that non-gaussian localisation effects only become apparent for $\tau > 1$, which we expect to be a result of the much shallower gradient in $V$ compared to the case of $D_R$ ($D_{R,max}/D_{R,min} = 100\cdot V_{max}/V_{min}$), and the resulting need for less-responsive particles for localisation effects to emerge.

We then evaluate the ability of our previously developed RNN to estimate $\hat{\tau_i}$ from $\hat{V_i}$ and $V_i(\vec{r})$ for $L/\langle V\rangle = 10$. We retain the architecture and pre-processing steps previously developed for analysing $D_R$ time series, but no longer take the common logarithm of the raw input data, as $V$ now only varies over 1 order of magnitude. We find that the RNN performs equally well, if not better, for particles moving in a landscape of $V.$ In particular, we see that the network appears to predict higher values of $\tau$ better than for $D_R$, which may result from the more direct influence that velocity has on the rABP dynamics and the resulting variations in $V_i(\vec{r})$. Nevertheless, we note here that, like for $D_R$, in this approach we require direct access to $\hat{V_i}$, the “true" particle velocity. In the absence of noise, i.e. where Brownian motion due to thermal fluctuations are negligible (i.e. $D_T, D_R \rightarrow 0$), this value could indeed be directly extracted from the instantaneous displacements of the particle, $\hat{V_i} = V_{0,i} = \sqrt{\Delta x_i^2 + \Delta y_i^2}/\Delta t$, where $\Delta x_i$ and $\Delta y_i$ denote the motion of the $i^{th}$ particle in x and y respectively over time-step $\Delta t$. \textit{Unfortunately}, in experiment one is typically plagued by the infernal nuisance that is noise \cite{Muinos-Landin2020,Tovey2023}, and it is at the very least necessary to account for Brownian effects on the particle dynamics. We therefore now investigate whether it is possible to extract $\tau$ from experimentally accessible parameters, i.e. from the trajectory of a rABP.

\begin{figure}
\centering
  \includegraphics[width=0.75\linewidth]{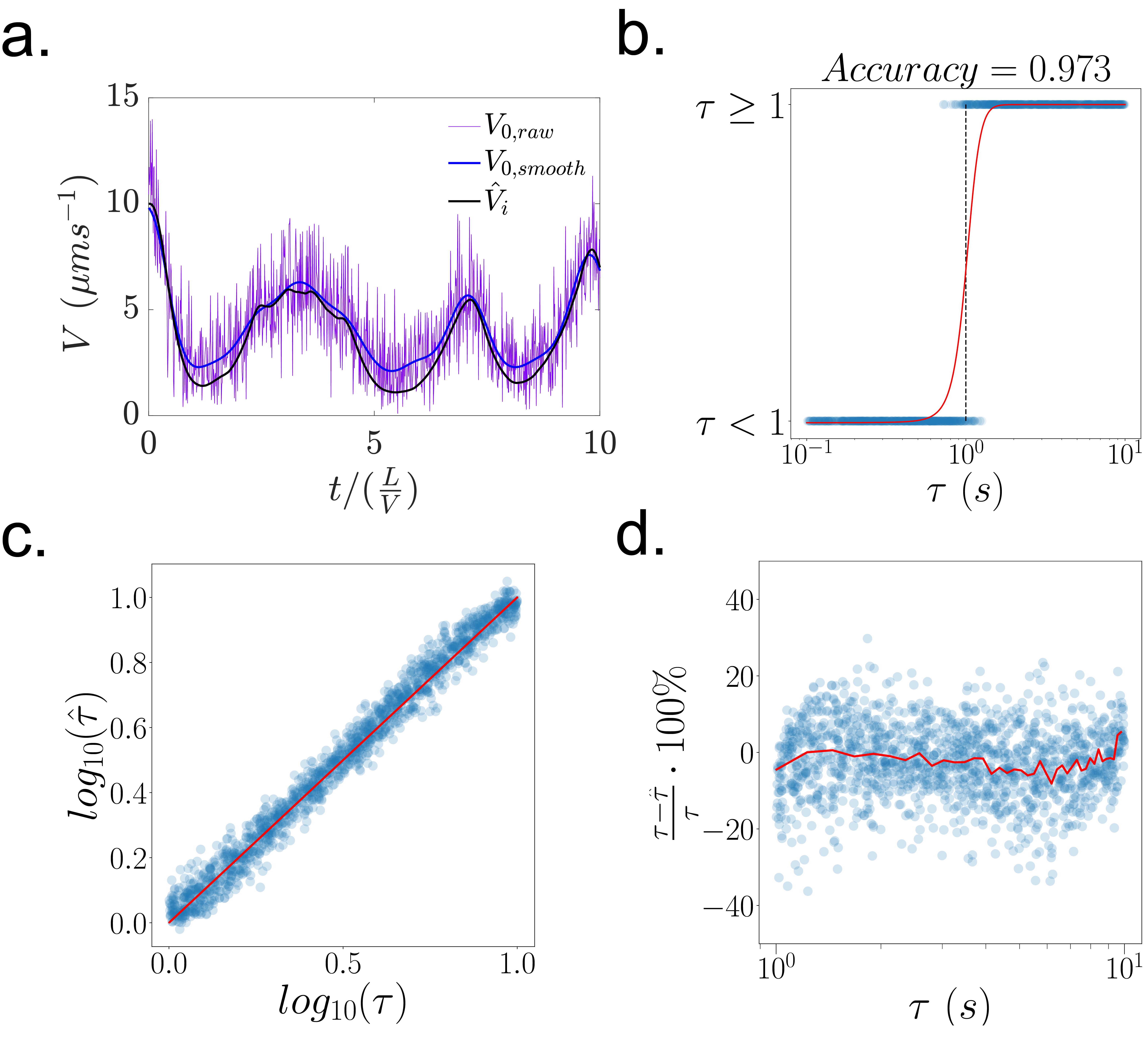}
  \caption{Investigating the ability of our GRU architecture to extract $\tau$ from particle trajectories, where the system time-scale is given by $L/\langle V \rangle = 10$. a) Translational and rotational diffusion, as well as sub-sampling the trajectories, introduce noise into the instantaneous velocity extracted from trajectories, $V_{0,raw}$ (here sub-sampled every 0.1 s). A Butterworth filter is applied to extract the de-noised signal $V_{0,smooth}$, which better approximates the underlying generating function $V(\vec{r})$. b) With the denoised time series $V_{0,smooth}$, a binary classifier can be constructed using the same GRU architecture which is able to classify with high accuracy between $\tau \geq 1$ and $\tau < 1$. The classifier was trained and evaluated on simulated data with 748 different $\tau$ values with 10 different particle realisations for each $\tau$. c) Fitting the response time $\tau$ from experimentally accessible data, for meaningful values of $\tau$ given $(V_{max}/V_{min})/(L/\langle V\rangle) = 1$ ($\tau \geq 1$). For smaller $\tau$ values, the response is effectively instantaneous given the small gradient in $V$. d) For $1\leq\tau\leq 10$, the normalised residuals obtained have minimal structure. We note that the effect of experimental “noise" widens the variance in $\hat{\tau}$. In c-d), we trained the network on 1000 realisations of $\tau$, log-spaced between 1 and 10.}
  \label{fig:Fig5}
\end{figure}

As before, we investigate the time series of rABPs generated for $L/\langle V\rangle = 10$. However, instead of using $\hat{V_i}$ directly from the simulations, we instead now attempt to extract an approximation from the instantaneous velocity of the particles, $V_{0,i} = \sqrt{\Delta x_i^2 + \Delta y_i^2}/\Delta t$. We note here that the determination of $V_{0,i}$ over two time steps requires that we do not use the final data point of $V_i(\vec{r})$. Immediately, the problems associated with working with experimental trajectories subject to noise and sampling effects become apparent (see Figure \ref{fig:Fig5}a). The values determined directly, $V_{0,raw}$, are too noisy to provide a faithful approximation of the underlying velocity $\hat{V_i}$, preventing model training. More specifically, the large differences between subsequent measurements of $V_{0,raw}$ is exacerbated by the differencing step required in pre-processing, creating a uninformative time series from which the RNN must learn. The issues associated with noise can be somewhat mitigated by increasing the sampling time, $\Delta t$, so as to increase the importance of the persistent motion $V$ relative to the thermal diffusion ($V\times\sqrt{D_T/\Delta t} > 1$), albeit at the cost of the time resolution. For $V_{min} = 1 \mu$s$^{-1}$ and $D_T = 0.137 \mu$m$^2$s$^{-1}$, as in the simulations, we therefore set $\Delta t = 0.1$s to ensure that the we study time-scales where the ballistic motion of the particles outweighs the contribution of diffusion. As we still require 1200 data points for analysis with the recurrent network, this doubles the effective experimental time. An equivalent active matter experiment would therefore require a trajectory of 2 minutes at 10 FPS, which is reasonable.

It can be seen in Figure \ref{fig:Fig5}a that this sub-sampling rate still provides a large degree of noise in the directly extracted $V_{0,raw}$. Further increasing $\Delta t$ was found not to significantly affect the issue, as the dynamics between time-steps then became increasingly important (with a continuously fluctuating $V(\vec{r})$). We therefore used standard signal processing techniques, namely using the transfer function coefficients of a 4th order Butterworth filter (using the \textit{MATLAB} in-built function \textit{butter} \cite{butterworth1930experimental}) with a zero-phase digital filter (using the \textit{MATLAB} in-built function \textit{filtfilt} \cite{Gustafsson1996}) applied to the noisy $V_{0,raw}$ to get the smoothed time series $V_{0,smooth}$. As we show in Figure \ref{fig:Fig5}a, the denoised signal well approximates the “true" velocity generating the motion observed, $\hat{V}_i$. 
Despite the ability to denoise the instantaneous velocity signal, and obtain reasonable approximations for $\hat{V}$, initial investigations indicated the inability to accurately estimate small values of $\tau$, which we attribute to the still prevalent effects of noise \cite{Muinos-Landin2020,Tovey2023}. Re-examining Figure \ref{fig:Fig4}a, we again note the threshold $\tau > 1$ to observe enhanced localisation effects for all $L/\langle V\rangle$. We suspect that the shallow gradient in $V$ ($(V_{max}/V_{min})/(L/V) = 1$s$^{-1}$) is hidden by the stochastic noise present in the simulation, such that for small $\tau$ - where the update in velocity is near-instantaneous - changes in velocity due to the landscape or because the effects of diffusion cannot be effectively decoupled. This would in turn explain why our RNNs are not able to satisfactorily train on real “experimental" data for $\tau < 1$, as the noise from thermal diffusion is of the same order as the particle acceleration resulting from the change in its position. 

To overcome this issue and still enable the prediction of reasonable $\tau$ values over which non-gaussian behaviours such as localisation are to be expected, we therefore propose a two-step approach to characterising the dynamics of rABPs. In the first step, we train the same RNN architecture, but now implement a binary classifier in the final layer using binary cross-entropy as the loss metric. More specifically, we label all values $\tau < 1$ as 0, and all values of $\tau \geq 1$ as 1, and are thereby able to obtain a very high accuracy on the test dataset of 0.973. Values of $\tau < 1$ are assumed to be effectively instantaneous, which is a reasonable assumption given for the Péclet number of the rABPs studied, where localisation only emerges for finite response times of $\tau > 1$. We then proceed to train our RNN architecture on the now experimentally relevant range of $1\leq\tau\leq 10$, and are able to obtain good estimates with minimal structures in the residuals (see Figures \ref{fig:Fig5}c-d). Notably, the width of the normalised errors has approximately doubled to 20\%, however, the effects of thermal noise on reducing the accuracy of neural networks is to be expected \cite{Muinos-Landin2020}, and we believe that the values for $\hat{\tau}$ we obtain with our approach are satisfactory.

\section{Conclusions}

We have demonstrated that our numerical model of responsive active Brownian particles (rABPs) can implement feedback strategies inspired from nature, and thereby introduce memory into the system. These memory effects in turn result in out-of-equilibrium dynamics such as enhanced localisation in more realistic spatial environments than those previously studied. Our approach can also be extended to include temporal fluctuations, enabling the modelling of e.g. active agents in complex dynamic environments such as travelling waves \cite{Lozano2019,Skoge2014}. Importantly, the continuous response of our rABPs provides us with time series of their dynamics, which not only enables the classification of their response via statistical analytical tools, but also opens the door to using recurrent neural networks (RNNs) to characterise their behaviour in a non-trivial manner. Specifically, RNNs provide us with insights into the physical parameters which govern the response of ABPs to their environment. By performing experiments with various microswimmer systems responding to externally imposed variations in their velocity (or other directed propulsion strategies, e.g. angular velocity), we expect that it will be possible to determine the responsiveness of different materials to their environment. Given the importance of memory effects for active materials, our proposed strategy could thus be used to guide the design of physically intelligent microrobots capable of autonomously detecting to their surroundings and performing pre-programmed tasks, helping realise yet outstanding challenges within the field.

\medskip
\textit{Acknowledgements} The authors would like to thank Dr. G Volpe, J Pineda, H. K. Moberg, and B. Midtvedt for insightful discussions and instruction. This project has received funding from the European Research Council (ERC) under the European Union’s Horizon 2020 research and innovation program grant agreement No. 101001514.
 
\medskip
\textit{Author Contribution Statement} - Author contributions are defined based on the CRediT (Contributor Roles Taxonomy). Conceptualization: M.R.B., F.G., L.I. Discussions: M.R.B., F.G., L.I. Formal Analysis: M.R.B. Funding acquisition: L.I. Investigation: M.R.B. Methodology: M.R.B. Software: M.R.B., F.G. Supervision: L.I. Validation: M.R.B. Visualization: M.R.B. Writing - original draft: M.R.B. Writing - review and editing: M.R.B., L.I.

\bibliographystyle{rsc}
\bibliography{responsive_ABP}

\end{document}